\documentclass{ws-procs9x6}

\newcommand {\bd}{\begin{displaymath}}
\newcommand {\ed}{\end{displaymath}}
\newcommand {\eq}{\begin{equation}}
\newcommand {\beq}{\begin{equation}}
\newcommand {\eeq}{\end{equation}}
\newcommand {\beqa}{\begin{eqnarray}}
\newcommand {\eeqa}{\end{eqnarray}}

\newcommand {\tr}{{\rm tr\,}}

\newcommand {\RR}{\mbox{\scriptsize R}}
\newcommand {\II}{\mbox{\scriptsize I}}

\newcommand {\trs}{\mbox{\scriptsize tr}}

\newcommand {\ee}{\mbox{e}}

\newcommand {\dd}{\mbox{d}}

\newcommand {\del}{\partial}

\newcommand {\defeq}{\stackrel{\rm def}{=}}

\newcommand {\vev} [1] {\langle #1 \rangle}
\newcommand{\id}{{1\!\!1}} 

\renewcommand{\theequation}{\thesection.\arabic{equation}}

\begin{document}

\title{A New Method for Simulating QCD\\
at Finite Density}

\author{Jun NISHIMURA}

\address{High Energy Accelerator Research Organization (KEK),\\
1-1 Oho, Tsukuba 305-0801, Japan\\
E-mail: jnishi@post.kek.jp}


\maketitle

\abstracts{
We propose a new method for simulating QCD at finite density,
where interesting phases such as the color superconductivity 
phase is conjectured to appear.
The method is based on a general factorization property of 
distribution functions of observables,
and it is therefore applicable to any system with 
a complex action. 
The so-called overlap problem is completely eliminated
by the use of constrained simulations.
We test this method in a Random Matrix Theory 
for finite density QCD, 
where we are able to reproduce the exact results for the 
quark number density. 
The achieved system size is large enough to 
extract the thermodynamic limit. Our results provide a 
clear understanding 
of how the expected first order phase transition
is induced by the imaginary part of the action.
We also discuss the noncommutativity
of the zero chemical potential limit ($\mu \rightarrow 0$)
and the thermodynamic limit, which is relevant to
recent Monte Carlo studies at small $\mu$.
}


\section{Introduction}
Recently there are a lot of activities in QCD 
at finite density, 
where interesting phases such as a superconducting phase
have been conjectured to appear \cite{bailin}.
At zero chemical potential
Monte Carlo simulations of lattice QCD
enable nonperturbative studies from first principles.
It is clearly desirable to extend such an approach 
to finite density and explore the phase diagram of QCD in the
$T$(temperature)-$\mu$(chemical potential) plane.
The main obstacle here is that the Euclidean action becomes complex
once the chemical potential is switched on.

Nevertheless QCD at finite density has been studied by various
approaches with exciting conjectures.
First there are perturbative studies which are valid 
in the $\mu \rightarrow \infty$ limit \cite{Son:1998uk,Schafer:1999jg}.
Refs.\ [\refcite{krishna}] and [\refcite{edward}]
use effective theories
with instanton-induced four-fermi interactions.
As for Monte Carlo studies two directions have been pursued so
far. One is to modify the model so that the action becomes
real. This includes changing the gauge group 
from SU(3) to SU(2) \cite{Kogut:2002cm},
and introducing a chemical potential with opposite signs for up and down
quarks \cite{Kogut:2002zg}.
%
The other direction is to explore the large $T$ and small
$\mu$ regime of lattice QCD, where the imaginary part of the action is
not very large 
\cite{Fodor:2001au,Allton:2002zi,deForcrand:2002ci,D'Elia:2002gd}. 
These studies already produced results relevant to heavy ion
collision experiments,
but more interesting physics will be uncovered 
if larger $\mu$ regime becomes accessible by simulations.

In Ref.\ [\refcite{Anagnostopoulos:2001yb}]
we have proposed 
a new method to simulate systems with a complex action,
which utilizes a simple factorization property of 
distribution functions of observables.
Since the property holds quite generally,
the approach can be applied to any system with a complex action.
The most important virtue of the method is that
it eliminates the so-called overlap
problem, which occurs in the standard re-weighting method.
Ultimately we hope that this method will enable us, among other things,
to explore the phase diagram of QCD at finite baryon density.

As a first step we test \cite{Ambjorn:2002pz}
the new approach in a Random Matrix Theory, 
which can be regarded as a schematic
model for QCD at finite baryon density \cite{Stephanov:1996ki}.
We also present preliminary results \cite{prelim}, which reveal certain
noncommutativity of the $\mu \rightarrow 0 $ limit and the
thermodynamic limit.


\section{Brute-force approach ---reweighting method---}
\label{old_approach}
\setcounter{equation}{0}
\renewcommand{\thefootnote}{\arabic{footnote}}

Suppose we want to study the model defined by the partition function
\beq
Z = \int \dd U \, \ee ^{-S_0 + i \, \Gamma}\ ,
\label{rmtdef2}
\eeq
where $S_0$ and $\Gamma$ are real.
Since the weight $\ee ^{-S_0 + i \, \Gamma}$ in (\ref{rmtdef2})
is not positive definite, 
we cannot regard it as a probability density.
Hence it seems difficult to apply the idea of standard Monte Carlo
simulations,
which reduces the problem of obtaining vacuum expectation values
(VEVs) to that of taking 
an average over an ensemble generated by the probability.
One way to proceed is to apply the reweighting method
and rewrite the VEV $\langle {\mathcal O} \rangle$ as
\beq
\left\langle {\mathcal O} \right\rangle
= \frac{\left\langle {\mathcal O} \, \ee ^{i \Gamma }
\right\rangle _{0}}
{\left\langle \ee ^{i \Gamma }
\right\rangle _{0}}  \ ,
\label{VEV}
\eeq
where the symbol $\langle \ \cdot \ \rangle _{0}$ denotes 
a VEV with respect to the {\em phase-quenched} 
partition function
\beq
Z_0
= \int \dd U \, \ee ^{-S_0}\ .
\label{absdef}
\eeq
Since the system (\ref{absdef}) has a positive definite weight,
the VEV $\langle \ \cdot \ \rangle _{0}$ can be evaluated by
standard Monte Carlo simulations.
However, the fluctuations of the phase $\Gamma $ in (\ref{VEV}) 
grows linearly with the system size $V$.
Due to huge cancellations,
both the denominator and the numerator of the r.h.s.\ of (\ref{VEV})
vanish as $\ee ^{-{\rm const.} V}$ as $V$ increases,
while the `observables' 
$\ee ^{i \Gamma }$ and ${\mathcal O} \ee ^{i \Gamma }$
are of O(1) for each configuration.
As a result, 
the number of configurations required to obtain the VEVs with
some fixed accuracy grows as $\ee ^{{\rm const.} V}$. 
This is the notorious `complex-action problem'.
Moreover when one simulates the phase-quenched model
(\ref{absdef}), one cannot sample efficiently the configurations
which are relevant to the calculation of 
the VEV $\langle {\mathcal O} \rangle$.
This is the so-called `overlap problem'.

\section{New approach --- factorization method ---}
\label{new_approach}
\setcounter{equation}{0}
\renewcommand{\thefootnote}{\arabic{footnote}}

In the factorization method proposed in 
Ref.\ [\refcite{Anagnostopoulos:2001yb}],
the fundamental objects are the distribution functions
(we assume the observable ${\mathcal O}$ to be real)
\beqa
\rho(x) &\defeq& \langle \delta (x - {\mathcal O}  )\rangle \\
\rho^{(0)} (x) &\defeq& \langle \delta (x - {\mathcal O} )\rangle_0
\eeqa
defined for the full model (\ref{rmtdef2})
and for the phase-quenched model (\ref{absdef}), respectively.
The important property of $\rho(x)$ is that it factorizes as
\beq
\rho (x) = \frac{1}{C} \, \rho^{(0)} (x) \, \varphi (x)   \ ,
\label{factorize}
\eeq
where the constant  $C$ is given by
$C \defeq \langle \ee ^{i \, \Gamma}\rangle _0$.
The `weight factor' $\varphi(x)$, which represents the effect of 
$\Gamma$, can be written as a VEV
\beq
\varphi (x) \defeq \langle \ee ^{i\, \Gamma} \rangle_{x}
\eeq
with respect to yet another partition function
\beq
Z(x) = \int \dd U \, \ee^{-S_0} \, \delta(x- {\mathcal O}) \ .
\label{cnstr_part}
\eeq
The $\delta$-function represents a constraint on the system.
In actual simulation we replace 
the $\delta$-function by a sharply peaked potential.
The distribution $\rho^{(0)} (x)$ for the phase-quenched model
can also be obtained from the same simulation.
Then the VEV $\langle {\mathcal O} \rangle$ can be obtained by
\beq
\langle {\mathcal O} \rangle
= \int \dd x \, x \, \rho (x)
=\frac
{\int \dd x \, x \, \rho^{(0)} (x) \, \varphi(x)}
{\int \dd x \,  \rho^{(0)} (x) \, \varphi(x)} \ ,
\eeq
where the overlap problem
is eliminated by forcing the simulation to sample the important
configurations by the constraint.
The knowledge of the weight factor $\varphi(x)$ is useful
because it tells us precisely which values of ${\mathcal O}$
are favored or disfavored by the effects of the oscillating phase.
Once a rough estimate of $\rho (x)$ is obtained, one may perform
multi-canonical simulations with an appropriate weight
(instead of simulating (\ref{cnstr_part}) for many $x$)
to sample relevant configurations more efficiently.
This has not yet been done, however.

\section{Random Matrix Theory for finite density QCD}
\label{RMTintro}
\setcounter{equation}{0}
\renewcommand{\thefootnote}{\arabic{footnote}}

The Random Matrix Theory we study is defined by the 
partition function
\beq
Z = \int \dd W \ee ^{- N \, \trs (W^\dag W)} \, \det D  \ ,
\label{rmtdef}
\eeq
where $W$ is a $N \times N$ complex matrix, and 
$D$ is a $2N \times 2N$ matrix given by
\beq
D = 
\left( 
\begin{array}{cc}
m & i W +  \mu  \\
i W^\dag  + \mu & m
\end{array}
\right) \ .
\label{defD}
\eeq
The parameters $m$ and $\mu$ correspond to the 
`quark mass' and the `chemical potential', respectively. 
The fermion determinant becomes complex for $\mu \neq 0$,
so we write it as
$\det D =  \ee ^ {i \Gamma}\,  | \det D |$.
The complex-action problem arises due to the phase $\Gamma$.
In what follows we consider the massless case ($m=0$) for simplicity
and focus on the `quark number density' defined by
\beq
\nu   =  \frac{1}{2N} \, \tr \, ( \gamma_4 D^{-1} ) \ ,
\mbox{~~~~~~~} \gamma_4 = \left( 
\begin{array}{cc}
0 & \id  \\
\id  & 0
\end{array}
\right) \ .
\eeq

The model was first solved in the large-$N$ limit \cite{Stephanov:1996ki},
and turned out to be solvable later even for finite $N$ 
\cite{Halasz:1997he}.
The partition function can be expressed as
\beq
Z(\mu , N) = \pi \ee^{\kappa} N ^{-(N+1)} \, N !
\left[ 1+ \frac{(-1)^{N+1}}{N !} \gamma (N+1 ,\kappa)
\right] \ ,
\label{partitionfn}
\eeq
where $\kappa = - N \mu^2$ and $\gamma(n,x)$ is the incomplete 
$\gamma$-function defined by
\beq
\gamma(n,x)=\int_0 ^x \ee^{-t} \, t^{n-1} \, \dd t \ .
\eeq
From this one obtains the VEV of the quark number density as
\beqa
\langle \nu \rangle
&=& \frac{1}{2N} \frac{\del}{\del \mu} \ln Z(\mu , N) \\
&=& - \mu \left[ 1 + \frac{\kappa ^N \ee^{-\kappa}}
{ (-1)^{N+1}N\! + \gamma(N+1 , \kappa)   }
\right] \ .
\label{finiteN}
\eeqa
Taking the large-$N$ limit, one obtains
\beq
\lim_{N\rightarrow \infty}
 \langle \nu \rangle =
\left\{ \begin{array}{ll}
- \mu  & \mbox{~~~for~}\mu < \mu _{\rm c}  \\
1/ \mu  & \mbox{~~~for~}\mu > \mu _{\rm c} \ ,
\end{array} 
\right. 
\label{nq_rmt}
\eeq
where $\mu_{\rm c}$ is the solution to
the equation $1 + \mu^2 + \ln (\mu^2) = 0$,
and its numerical value is given by $\mu_{\rm c} =0.527\cdots$.
We find that the quark number density $\langle \nu \rangle$
has a discontinuity at $\mu = \mu_{\rm c}$.
Thus the schematic model reproduces qualitatively 
the first order phase transition expected to occur
in `real' QCD at nonzero baryon density.


The phase-quenched model defined by the partition function
\beq
Z_0 = \int \dd W \ee ^{- N \, \trs (W^\dag W)} \, |\det D | 
\label{phasequencheddef}
\eeq
can be solved in the large $N$ limit \cite{Stephanov:1996ki} 
and one obtains
\beq
\lim_{N\rightarrow \infty}
\langle \nu \rangle_0 = 
\left\{ \begin{array}{ll}
\mu  & \mbox{~~~for~}\mu < 1  \\
1/ \mu   & \mbox{~~~for~}\mu > 1 \ ,
\end{array} 
\right. 
\label{nq_abs}
\eeq
which is a continuous function
of the chemical potential $\mu$ unlike (\ref{nq_rmt}).
Thus the first order phase transition
in the full model (\ref{rmtdef})
occurs precisely due to the imaginary part $\Gamma$ of the action.
This model therefore provides a nice testing ground for simulation
techniques for finite density QCD \cite{Stephanov:1996ki}.

\section{Testing the factorization method in the RMT}
\label{problem}
\setcounter{equation}{0}
\renewcommand{\thefootnote}{\arabic{footnote}}

Since $\nu$ is complex for each configuration,
we decompose it into the real and imaginary parts 
as $\nu = \nu_{\rm R} + i \nu_{\rm I}$
and calculate $\langle \nu_{\rm R} \rangle$ and 
$\langle \nu_{\rm I} \rangle$
by the factorization method
($\langle \nu_{\rm I} \rangle$ is purely imaginary).
We introduce the distribution functions
for $\nu_{\rm R}$ and $\nu_{\rm I}$ separately as
\beqa
\rho_i(x) &\defeq& \langle \delta (x - \nu_i  )\rangle 
~~~~~~i={\rm R,I}  \ , \\
\rho^{(0)}_i (x) &\defeq& \langle \delta (x - \nu_i )\rangle_0
~~~~~~i={\rm R,I}  \ .
\eeqa
The factorization holds for both $\rho_{\rm R}(x)$ and
$\rho_{\rm I}(x)$ as
\beq
\rho_i(x) = \frac{1}{C} \, \rho^{(0)}_i (x) \, \varphi _i(x)  
~~~~~~i={\rm R,\,I} \ ,
\label{factorize2}
\eeq
where the constant  $C$ is given by
$C \defeq \langle \ee ^{i \, \Gamma}\rangle _0$.
The weight factors $\varphi _i(x)$ can be written as a VEV
\beq
\varphi_i(x) \defeq \langle \ee ^{i\, \Gamma} \rangle_{i,\, x}
~~~~~~i={\rm R,\,I}
\eeq
with respect to the constrained phase-quenched model
\beq
Z_i(x) = \int \dd W \, 
\ee ^{- N \, \trs (W^\dag W)} \, |\det D |
 \, \delta(x-\nu_i) 
~~~~~~i={\rm R,\,I} \ .
\label{cnstr_part2}
\eeq


Under the transformation $W \mapsto - W$, the
Gaussian action is invariant, whereas the fermion determinant 
$\det D$ as well as the quark number density $\nu$ becomes
complex conjugate.
Due to this symmetry, we have
\beqa
\varphi_{\rm R} (x)^* &=& \varphi_{\rm R}  (x)  \ , \\
\varphi_{\rm I} (x) ^* &=& \varphi_{\rm I} (-x)  \ , \\
\rho^{(0)} _{\rm I} (-x) &=& \rho^{(0)} _{\rm I} (x)  \ .
\eeqa
Using these properties, we arrive at
\beqa
\label{simpleformula0}
\langle \nu_{\rm R} \rangle &=&
\frac{1}{C} 
 \int _{-\infty} ^{\infty}
 \dd x \, x \, \rho_{\rm R}^{(0)} (x) \, w_{\rm R}(x)  \ , \\
\langle \nu_{\rm I} \rangle &=&
  \frac{2 \, i}{C} 
\int  _{0} ^{\infty}
\dd x \, x \, \rho ^{(0)}_{\rm I} (x) \, w_{\rm I} (x) \ , 
\label{simpleformula}    \\
C &=& \int _{-\infty} ^{\infty} \dd x 
\, \rho^{(0)} _{\rm R} (x) \,  w _{\rm R} (x) \ ,
\label{C_new}
\eeqa
where the weight factors $w_i(x)$ are defined by
\beqa
\label{wRcos}
w_{\rm R}  (x) &\defeq& \langle  \cos \Gamma \rangle_{\RR , x }  \\
w_{\rm I}(x) &\defeq& \langle \sin \Gamma\rangle_{\II , x} 
= - w_{\rm I}(-x) \ .
\label{wIsin}
\eeqa

\begin{table}[ph]
\tbl{Results of the analysis of $\langle \nu \rangle $
described in the text.
Statistical errors computed by the jackknife method are shown.
The last column represents the exact result 
for $\vev{\nu}$ at each $\mu$ and $N$.
For $\mu=0.2$ the exact result is $\vev{\nu}=-0.2$ 
with an accuracy better than $1$ part in $10^{-9}$. }
{\footnotesize
\begin{tabular}{|c | c | c | c | c | c | }
\hline
$\mu$&$N$ &$\vev{\nu_{\rm R}}$&$i \, \vev{\nu_{\rm I}}
$&$\vev{\nu}$&$\vev{\nu}$~(exact)\\
\hline
 0.2 & 8  & 0.0056(6) & -0.1970(5)  & -0.1915(7) & -0.20000\ldots \\
 0.2 & 16 & 0.0060(4) & -0.1905(13) & -0.1845(13)& -0.20000\ldots \\
 0.2 & 24 & 0.0076(9) & -0.1972(14) & -0.1896(17)& -0.20000\ldots  \\
 0.2 & 32 & 0.0021(8) & -0.1947(19) & -0.1927(25)& -0.20000\ldots  \\
 0.2 & 48 & 0.0086(37)& -0.2086(54) & -0.2000(88)& -0.20000\ldots  \\
\hline
 1.0 &  8 & 0.8617(10)&  0.1981(13) &  1.0598(12)&1.066501$\ldots$\\
 1.0 & 16 & 0.8936(2) &  0.1353(6)  &  1.0289(5) &1.032240$\ldots$\\
 1.0 & 32 & 0.9207(1) &  0.0945(2)  &  1.0152(3) &1.015871$\ldots$ \\
\hline
\end{tabular}
\label{t:1} }
\end{table}

In Table\ \ref{t:1} we show our results for two values of $\mu$, 
$\mu = 0.2$ and $\mu = 1.0$, which are on opposite  
sides of the first order 
phase transition point $\mu = \mu_{\rm c}=0.527 \cdots$.
They are in good agreement with the exact results,
and the achieved values of $N$ are 
large enough to extract the large $N$ limit.

\begin{figure}[htbp]
  \begin{center}
    \includegraphics[height=7cm]{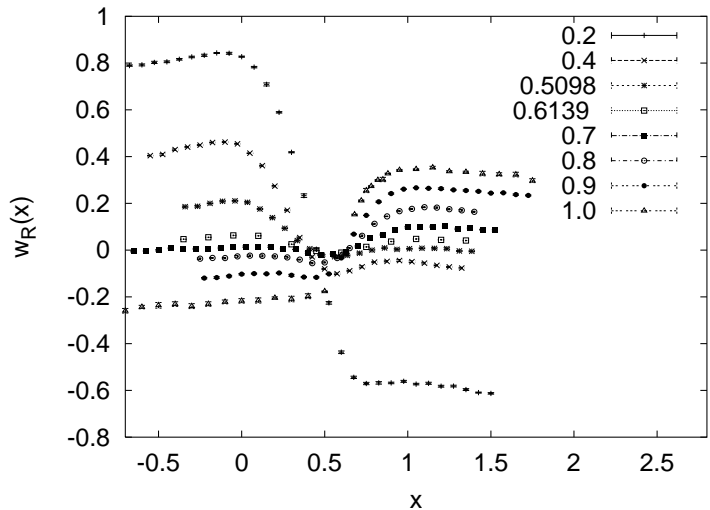}
    \caption{The weight factor $w_{\rm R}(x)$ is plotted 
for $N=8$ at various $\mu$.
}
    \label{fig:wR_N8}
  \end{center}
\end{figure}
\begin{figure}[htbp]
  \begin{center}
    \includegraphics[height=7cm]{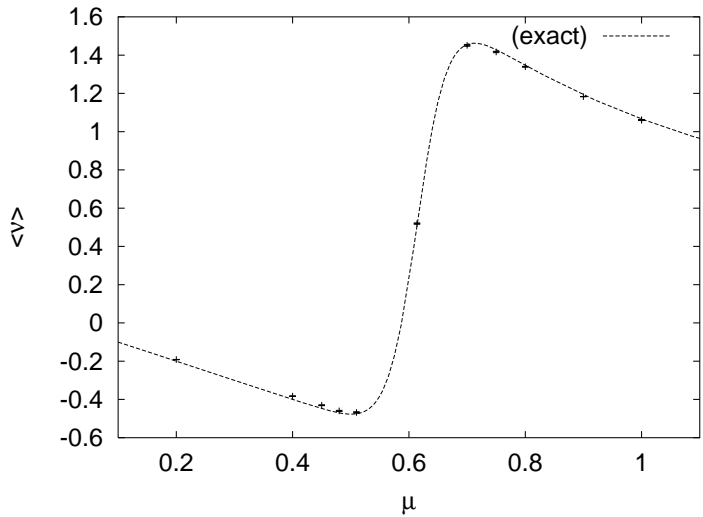}
    \caption{
The VEV $\vev{\nu}$ obtained by the factorization
method is plotted against $\mu$ for $N=8$ including the critical regime. 
Statistical errors computed by the jackknife method are also shown.
The dashed line represents the exact result 
(\protect\ref{finiteN}) for $\vev{\nu}$ at $N=8$.}
    \label{fig:crit_N8}
  \end{center}
\end{figure}

In Fig.\ \ref{fig:wR_N8}
we plot $w_{\rm R}(x)$
for $N=8$ at various $\mu$.
It is interesting that
the $w_{\rm R}(x)$ changes from positive to negative for 
$\mu < \mu_{\rm c}$,
but it changes from negative to positive for $\mu > \mu_{\rm c}$.
(Similarly $w_{\rm I}(x)$ is positive at $x>0$ for $\mu < \mu_{\rm c}$,
but it is negative at $x>0$ for $\mu > \mu_{\rm c}$.)
Thus the behavior of $w_i (x)$ changes drastically as 
the chemical potential $\mu$ crosses its critical value $\mu_{\rm c}$.
These results provide a clear understanding of 
how the first order phase transition occurs
due to the effects of $\Gamma$.
Fig.\ \ref{fig:crit_N8} shows
the results for $\langle \nu \rangle$ obtained by the factorization
method for $N=8$ at various $\mu$, 
which nicely reproduce the gap developing at the critical point.

\section{Noncommutativity of $\mu \rightarrow 0 $ and 
$N \rightarrow\infty$}
\label{noncomm}
\setcounter{equation}{0}

In this Section we discuss
the noncommutativity of 
the two limits, $\mu \rightarrow 0$ and $N \rightarrow\infty$.
The absence of such noncommutativity is implicitly assumed in most 
of the recent approaches used in simulating finite density QCD
at small $\mu$. This includes the multi-parameter reweighting
approach \cite{Fodor:2001au},
the Taylor expansion approach \cite{Allton:2002zi},  
and the imaginary $\mu$ approach \cite{deForcrand:2002ci,D'Elia:2002gd}.
In all these works one restricts oneself to the small $\mu$
regime where the fluctuation of the phase is still under control.

In fact the noncommutativity can be readily seen from
the exact result (\ref{partitionfn})
for the partition function (\ref{rmtdef}).
%
The phase of the determinant vanishes at $\mu = 0$
for finite $N$, and one obtains a
nonzero result for the partition function $Z=1$ in the large $N$ limit.
On the other hand, the oscillation of the phase
becomes pronounced at sufficiently large $N$ even for small but finite $\mu$,
and as a result one obtains $Z=0$ in the large $N$ limit as far as
$\mu$ is kept finite.
This implies in particular that the free energy 
\beq
f(\mu) = - \lim_{N\rightarrow \infty}
\frac{1}{N^2} \ln Z(\mu , N)
\eeq
has a discontinuity at $\mu = 0$ as
\beq
\lim _{\mu \rightarrow 0} f(\mu) > f(0) = 0 \ .
\eeq
We expect that, in general,
the free energy of 
a system with a complex action
has a discontinuity at a point in the parameter space where
the imaginary part of the action vanishes identically.

With the factorization method 
we can take the two limits $\mu \rightarrow 0 $,
$N \rightarrow \infty$
in different orders and compare the results.
As we will see, we do observe the noncommutativity in various ways.
On the other hand, we know from the exact result 
(\ref{finiteN}) that the VEV $\langle \nu \rangle$
does not have the noncommutativity.
In the factorization method $\langle \nu \rangle
=\langle \nu_{\rm R} \rangle +i\langle \nu_{\rm I} \rangle$
is calculated by the formulae
(\ref{simpleformula0}), (\ref{simpleformula}) and (\ref{C_new}).
In fact the functions $w_{\rm R}(x)$, $w_{\rm I}(x)$ and 
$\rho^{(0)}_{\rm R}(x)$ have the noncommutativity, but
these effects cancel each other in the end results for 
$\langle \nu_{\rm R} \rangle$ and $\langle \nu_{\rm I} \rangle$.
In what follows we present preliminary results relevant to 
$\langle \nu_{\rm R} \rangle$,
but similar results are obtained for $\langle \nu_{\rm I} \rangle$ 
as well \cite{prelim}.

\begin{figure}[htbp]
  \begin{center}
    \includegraphics[height=7cm]{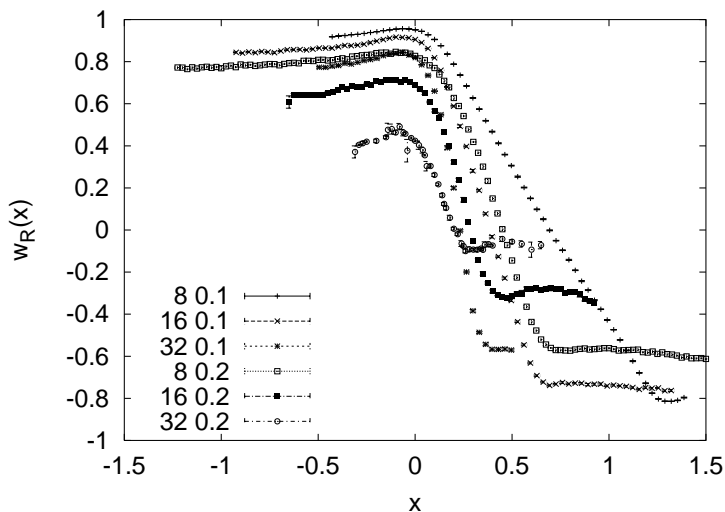}
    \caption{
The weight factor $w_{\rm R}(x)$ is plotted
for $\mu = 0.1$ and 0.2 at $N= 8,16,32$.
}
    \label{fig:rhoR}
  \end{center}
\end{figure}

Let us first look at the weight factor $w_{\rm R}(x)$,
which has the noncommutativity similar to the partition function.
At $\mu = 0$ one obtains $w_{\rm R}(x)\equiv 1 $ for any $N$, 
whereas in the large $N$ limit one obtains $w_{\rm R}(x) \equiv 0$ for
any $\mu$.
In Fig.\ \ref{fig:rhoR} we plot $w_{\rm R}(x)$ for $\mu = 0.1$
and $\mu = 0.2$ at $N=8,16,32$.
It shows clearly that the behavior of $w_{\rm R}(x)$ depends much
on the order of the two limits 
$\mu \rightarrow 0$, $N\rightarrow \infty$.

\begin{figure}[htbp]
  \begin{center}
    \includegraphics[height=7cm]{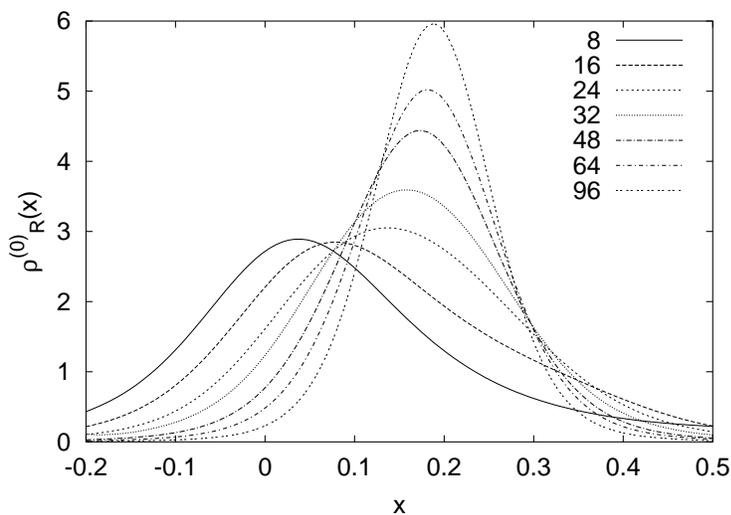}
    \caption{The function $\rho _{\rm R} ^{(0)} (x)$
is plotted
for $\mu = 0.2$ at various $N$.
}
    \label{fig:rho0R}
  \end{center}
\end{figure}

\begin{figure}[htbp]
  \begin{center}
    \includegraphics[height=7cm]{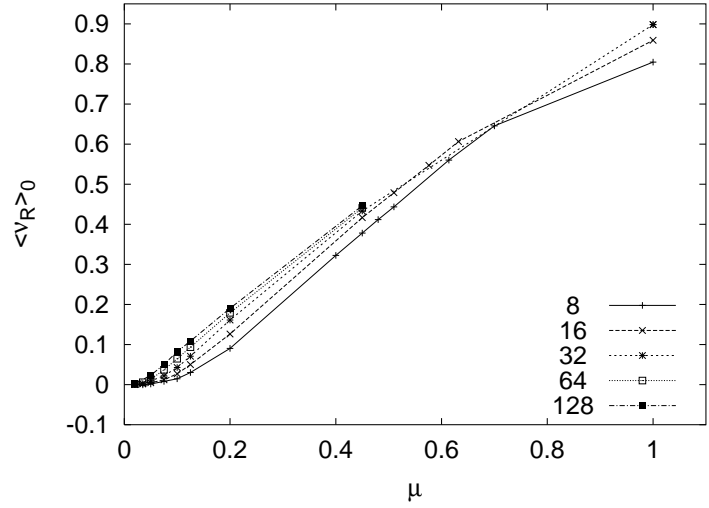}
    \caption{
The VEV $\langle \nu_{\rm R} \rangle _0$ 
is plotted against $\mu$ for various $N$.
}
    \label{fig:nuR0}
  \end{center}
\end{figure}

Let us next turn to $\rho _{\rm R} ^{(0)} (x)$.
In Fig.\ \ref{fig:rho0R} we plot it for various $N$ at $\mu = 0.2$.
At small $N$ the distribution is peaked near the origin
and the dependence on $N$ is small.
At sufficiently large $N$ the peak moves to $x \sim \mu$
and starts to grow, 
which is consistent with the large $N$ result 
$\vev{\nu_{\rm R}}_0 = \mu$.
Empirically we find that the transition occurs at
\beq
N_{\rm c} = \frac{0.25\sim 0.3}{\mu^2} \ .
\eeq
Thus the distribution $\rho _{\rm R} ^{(0)} (x)$ for 
the phase-quenched model also depends much on the order of the two limits
$\mu \rightarrow 0$, $N \rightarrow \infty$.
This can be seen more clearly in Fig.\ \ref{fig:nuR0}, where 
we plot the VEV $\langle \nu_{\rm R}
\rangle _0$ against $\mu$ for various $N$.
In particular the derivative
$ \frac{\del}{\del \mu} \langle \nu_{\rm R}\rangle _0$
becomes 0 if one takes the $\mu \rightarrow 0$ limit first,
but it becomes 1 if one takes the $N \rightarrow \infty$ limit first.

\begin{figure}[htbp]
  \begin{center}
    \includegraphics[height=7cm]{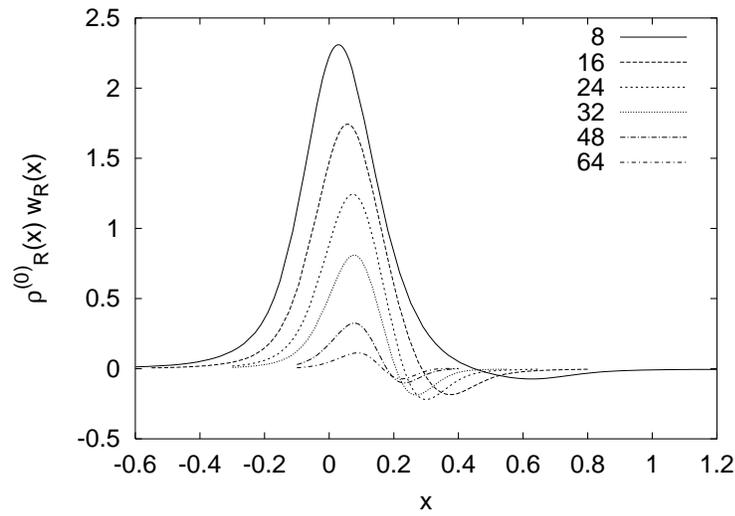}
    \caption{
The product $\rho^{(0)}_{\rm R}(x) w_{\rm R}(x)$,
which gives the unnormalized distribution 
for $\nu _{\rm R}$ in the full model,
is plotted for $\mu = 0.2$ at various $N$.
}
    \label{fig:rhoR_full}
  \end{center}
\end{figure}

The product $\rho^{(0)}_{\rm R}(x) \, w_{\rm R}(x)$ 
gives the unnormalized distribution for $\nu _{\rm R}$
in the full model,
which we plot in Fig.\ \ref{fig:rhoR_full}.
The distribution itself, even after appropriate normalization, 
has the noncommutativity, but 
the VEV $\langle \nu_{\rm R} \rangle$ calculated by the formula
(\ref{simpleformula0}) is always closed to zero (see Table \ref{t:1}).
The reason depends on the order of the two limits. 
If we take the $N\rightarrow \infty$ limit first, 
the positive and negative regions of $\rho_{\rm R}(x)$ cancel each other
in the calculation of the first moment.
If we consider small $\mu$ first, the distribution
is peaked around the origin, which makes the first moment close to zero.
Thus the noncommutativity cancels in the end result
for the VEV $\langle \nu_{\rm R} \rangle$.

\section{Concluding remarks}
\label{Summary}
\setcounter{equation}{0}


The factorization method has been applied also to other systems with
complex actions.
In the original paper \cite{Anagnostopoulos:2001yb},
it was used to study the dynamical generation of space time 
in superstring theory based on its matrix model
formulation \cite{IKKT}.
In this case the weight factors turned out to be positive definite,
which enabled us to use their scaling property
to make an extrapolation to larger system size.
The method \cite{Azcoiti:2002vk}
proposed for simulating $\theta$-vacuum like systems
can be regarded as a special case of the factorization method.
Promising results are obtained in 
the 2d CP$^3$ model etc.






\section*{Acknowledgments}
This work is partially 
supported by Grant-in-Aid for 
Scientific Research (No.\ 14740163) 
from the Ministry of Education, 
Culture, Sports, Science and Technology.

\end{document}